\documentclass[12pt, preprint]{aastex}
\usepackage{bm}
\begin{document}
\title{Dynamical Oscillations and Glitches in AXPs}
\author{David Eichler \altaffilmark{1},
Rashid Shaisultanov\altaffilmark{2}} \altaffiltext{1}{Physics
Department, Ben-Gurion University, Be'er-Sheva 84105, Israel;
eichler@bgu.ac.il} \altaffiltext{2}{Physics Department, Ben-Gurion
University, Be'er-Sheva 84105, Israel; rashids@bgu.ac.il}

\begin{abstract}
Recently reported observations of magnetar glitches and coincident X-ray long term brightening events establish enough of a data base to indicate that  the brightening events are accompanied by glitches at their outset, and that they are probably triggered by the same event that triggers the glitch. We suggest, on the basis of various observational clues, that a) these are caused by energy releases at depths below 100 meters, and b) the unpinning is due to global  mechanical motion triggered by the energy release, not by heat.  Because mechanical triggering of a glitch requires less energy than a detectable long term X-ray brightening, the latter does not necessarily accompany every glitch, but it is predicted, when it does occur, to be heralded by a glitch.
Crustal oscillation associated with the mechanical energy release may  cause short ($\ll 10^3$s) flares.

\end{abstract}

\section{Introduction}
A growing
number of X-ray brightening events - transient X-ray outputs of order $10^{41}$ ergs or more that fade out over several months - have been recorded  by RXTE monitoring of the pulsed component of AXP 1E 1048.1-5937 (Dib, Kaspi and Gavriil [DKG], 2009).  They appear to be reasonably quiet in impulsive flaring, unlike the case of 1E 2259+586 (Woods et al, 2004), which showed $\gamma$-ray flaring and X-ray brightening both within 12 hours of the inferred time of the glitch. Significantly, some of the  brightening events have a measurable rise time of order several weeks and {\it all} are accompanied by a glitch that,  to within the
accuracy of the observational gaps (days to weeks), occurs at or before the onset of the X-ray rise (DKG 2009).  Only a fraction ($\sim 1/3$) of glitches, on the other hand, are accompanied by detectable pulsed
brightening events (DKG 2008, 2009).

Quasi-periodic oscillations (QPO) in the hyperflares of soft
gamma-ray repeaters (Israel et al 2005; Strohmayer and Watts [SW], 2005;
Watts and Strohmayer, 2006; SW 2006)
may provide us much information about the dynamical response of magnetar crusts to magnetically induced disturbances. For one thing, they suggest that the crust remains solid or "breaks", rather than  plastically deforms or melts, when subjected to large magnetic stress. The higher frequencies, $\nu \gtrsim 10^2$ hz where the period of the wave is smaller than the crossing time of a torsional wave across the magnetar,
 suggest (as noted in the discovery papers) high $l$ ($l \sim 10$)  torsional
oscillations of the crust  with $n=0$, and perhaps the very highest frequencies involving $n=1$.\footnote{ $l $ is the polynomial index for spherical harmonics. n is the number of nodes in the radial dependence.}
 The predominance of n=0 modes, those in which all of the strata  of the crust move together, imply that the trigger event is accompanied by good, prompt communication from top to bottom of the crust. It also suggests that the available energy density, which is kinetic, may be more stratified than  magnetic energy density, i.e. it may deposit more heat in  deep strata than in shallow ones. It is these properties of the crustal motion that we invoke below  in order to explain the relation of glitches to X-ray brightening events.

The observed QPO amplitudes set a lower limit on the mechanical energy liberated in the crust by the impulsive events. Given that the luminosity variations of the QPO modes can be as high as 10 or 20
percent (SW 2006), one suspects that the amplitude is probably near the
limiting strain that the crust can withstand.
 Recently Timokhin, Eichler and Lyubarsky (2007)[TEL07]
 proposed a model for
the quasiperiodic component of magnetar emission during the tail
phase of giant flares. In the model, the magnetospheric currents are
modulated by torsional motion of the surface, which introduces a
perturbation in that current, and this translates into a modulation
of the optical depth of the magnetosphere to cyclotron scattering.
The authors calculated that the amplitude A of neutron star (NS)
surface oscillation should be about $ 1\% $ of the NS radius
$R_{NS}$ in order to produce the observed QPO amplitudes of $\sim 10$\%, assuming that the currents are carried by a singly charged
plasma. For a high $l$ mode ($l \sim 10$), this implies a shear strain somewhat
larger than $0.01$.  \footnote{
 Horowitz and Kadau (2009) concluded from  molecular
dynamics simulations  that even without a magnetic field, the breaking
strain may be around 0.1. In addition, the large magnetic field of
the magnetar crust probably allows an even greater strain.} The amplitude can
be lower if the pair multiplicity, the number of pairs per charge
carrier, is larger.

TEL07 noted that the implied mechanical energy in the
oscillations, nearly $10^{45}(A_i/0.01R_{NS})^2$ ergs per mode i, would by itself  be enough to power observable crustal heating, especially considering that many  modes  are probably excited ($\sim \sum(2l+1)\sim l_{max}^2$ if most or  all modes through $l_{max}$ are excited). QPO
oscillations  have thus far not been reported from any X-ray
heating event other than the two giant flares. It is clear that the energy is more than enough to power the X-ray emission that is contained in the observed brightening events,  typically $\sim 10^{41}$ to $10^{43}$ ergs, even if most of the heat is lost to the core. Heat released in the lower crust tends to flow into the core, and it is possible that observed large X-ray brightening events have QPOs with amplitudes as high as $10^{-2}$, which, as shown below, is far more than enough to cause glitches.

Glitches, sudden jumps in rotation rate, typically lie in the range  $\Delta \nu/\nu \simeq
10^{-10}$ to $10^{-5}$. Magnetar glitches tend to cluster near the
larger end of this range
(DKG 2008).
They are explained by unpinning of vortices from nuclei in the lower crust. The area
density of the vortices is related to rotation rate by $n= 2\Omega
m_{n}/\pi \hbar =10^{3}\Omega$ lines per $\mathrm{cm^{2}}$, where $m_{n}$ is
neutron mass and $\Omega$ is the angular velocity. At densities
$3\times 10^{13}\leq \rho \leq 2\times 10^{14}$ $\mathrm{g/cm^{3}}$,
 vortices may be pinned to the crust.
 This allows the crust to slow down
without a matching slowdown in the neutron superfluid.  Glitches can
happen when
 the Magnus force
exceeds the pinning force per length $f_{p}$, thus unpinning the vortices. Once the vortices are unpinned, dissipative effects (interaction between the two superposed components) transfer angular momentum from the
superfluid to the crust.

 The
adjustment of the crustal spin frequency at the time of the glitch
is limited by the relative velocity difference to which the crust and superfluid have
equilibrated hitherto, which is in turn limited by steady vortex
creep. The rate of vortex creep strongly increases with temperature,
since the creep proceeds by the tunneling of the vortices through
potential barriers. For this reason, it is widely believed (e.g. DKG 2008 and references therein)
 that adolescent neutron stars such as the Vela pulsar,  which have cooler crusts, are more resistant to vortex creep, allow greater angular velocity differences between superfluid and crust, and have larger glitches than younger ones, such as the Crab pulsar.

\section{Motivation for a Mechanical Mechanism for Glitch Plus Heating}

Given the above, DKG (2008) have argued that AXPs, as a class,
would thus be expected to have glitches of smaller frequency jumps than
radio pulsars, given that AXPs are much younger and hotter at the
surface than most radio pulsars. This, they note, is not the case.
Here  we emphasize that the fact that AXPs are hot at the surfaces does not mean
that they are hot at the base of their crusts, the depth at which
vortex pinning occurs. In fact, the long term declining X-ray afterglows of
SGRs 1627-41 and 1900+14 (Kouveliotou et al, 2003, Eichler  et al,
2006), if interpreted as the surface emission from a cooling crust
following transient heating events, implies that the cores must be
cool ($T \ll 10^8K$), otherwise the lower crust could not cool rapidly by dumping heat into the core,  and afterglow would last even
longer. The relatively low temperature assumed by these cores and lower crusts may
be attributed to cooling by the direct URCA process
(Lattimer 1991, Gnedin et al. (2001), the relatively young ages of AXPs notwithstanding, and the efficiency of heat conduction at temperatures below $10^8$ (e.g. Gnedin, Yakovlev and Potekhin, 2001).

 Link and Epstein (1996), and Larson and Link (2002) have proposed that
sudden heat release could release vorticies and thus trigger glitches. This would
be consistent with the observation (DKG 2008)
that AXP glitches frequently coincide with transient increases in X-ray
emission. This proposal also seems predicated on the assumption that the lower crust was cool enough prior to the episodic heating.

Invoking cool cores for AXPs would enable the lowest layers of crust, which are in close contact with the core, to cool quickly if and when heated.  This would account for several
different curiosities  concerning magnetar glitches and X-ray
enhancements, (long term decline of persistent emission, large glitches) and might imply that heating of the crust could trigger a glitch. However, there are additional constraints.
First,  many glitches ("silent" glitches) are unaccompanied by such
radiative enhancements (Dib, Kaspi and Gavriil, 2009). This would suggest that, if glitches are
directly triggered by crustal heating, then in many cases this heating would necessarily be well below the surface. Such heat could eventually make its way to the surface in observable amounts part of the time, while not in others. This is corroborated by the resolved rise of some X-ray brightening events, where the  maximum brightness is reached only after several weeks.

The coincidence of the glitches with onset of X-ray brightening even to within several weeks - and, in one of four cases, to within 12 hours (Woods et al., 2004) - suggests that the communication between the top and bottom of the crust that enables the coincidence is much faster than the thermal conduction timescale (years). Any thermal model for the unpinning is thus challenged by the near  temporal coincidence of glitches and onsets of X-ray brightening events.

 A torsional stress emanating from the magnetosphere might impart torsional strain to the crust, but it is not clear that the strain would store enough energy in the bottom of the crust to trigger any significant heating there once the crust yields to the stress, as this yield is most likely to happen near the top of the crust, where the material is weaker.   The gradual rise of the X-ray brightening, the absence of  $\gamma$-ray flaring in the magnetosphere, and the "silent" glitches all pose additional challenge to models where the primary energy dissipation lies in the magnetosphere or the uppermost layers of the crust.

Motivated by the above, we suggest that the unpinning
 is {\it mechanically} induced rather than
thermally induced. We point out in what follows that even modest amplitude torsional oscillations, such as were observed after giant flares, create an extremely large Magnus force so that vortices are unpinned. This does not preclude other phenomena (e.g. heat release) that might also unpin vorticies in the same event, but it guarantees that the glitch is triggered by the same dynamical event that triggers the X-ray brightening.

Mechanical oscillations, possibly including the motion that initializes them, also have the distinguishing feature that the mechanical energy density in the crust is proportional to mass density, as opposed the magnetic energy, which is probably weakly dependent on density. Thus, whereas direct magnetic energy dissipation into heat  would probably produce the largest initial temperatures near the surface, where the energy per unit mass is the largest, mechanical oscillations allow the surface temperature to reach a maximum somewhat later than the onset of the flare, even if the original source of the oscillation is magnetic energy. This is significant, because delays between onset and maximum,  of order several weeks are in some cases observed  (DKG 2009). In one such brightening event, the rise was reportedly 2.6 years (Dib et al 2007, Gonzalez et al 2009). This should be compared to the X-ray emission after giant flares which reaches its maximum much more quickly and then decays over several weeks. The latter timescale corresponds to the thermal diffusion of the outer 200 meters or so of the crust (Lyubarsky, Eichler and Thompson 2002), and the X-ray light curve in the six weeks  following the Aug. 27 event of SGR 1900+14 was fit assuming constant heat density input due to magnetic dissipation.

We suggest that the observed rise of several weeks may correspond to the time for heat to diffuse up to the surface after being generated  by dissipation of mechanical energy comfortably ($\sim 10^2$m or more) below the surface, though still in the outer crust. (Alternatively, one could simply postulate that the magnetic energy release that powers the X-ray brightening just happens to occur several hundred meters into the crust and below, but this would beg the question of why the release happens over most of the crust to the exclusion of just the outer 100 to 200 meters.) Invoking torsional oscillations admits a possible answer to this question inasmuch as it predicts an energy deposition that is proportional to the product of local dissipation rate and mass density, as opposed to the magnetic energy density, as a function of depth, even if the magnetic field is the principal source of energy. 

The long decays of the brightening events are  consistent with  deep crustal heating. While long decays may also be consistent with dissipation of small scale magnetospheric currents, the rise times of several weeks, the occasional anti-correlation of pulse fraction to brightness (Tiengo et al, 2005; Tam et al, 2008), and the absence of $\gamma$-rays for the three AXP 1E 1048.1-5937 events challenge the latter explanation.

\section{Mechanical Unpinning}

An oscillation amplitude of $A  = 0.01 R_{NS}$ would mean that
the velocity of crustal oscillations  is  $\sim 2\pi \times 10^{6} $
cm/s for frequency $\nu \sim 100$ Hz. We can expect that the
amplitude of n=0 (nodeless) toroidal oscillations
 does not change significantly with the radius in the
crust, see e.g. Lapwood and Usami (1981).

When $n \neq 0$ the amplitude inside the crust could become even
larger than on the surface of the crust ( Lapwood and Usami, 1981).
If, as seems likely, many modes are excited, the root mean square of the crustal velocity may
be an order of magnitude higher than for any given mode. Altogether,
we conclude that for oscillation amplitudes of $A_{-2}10^{-2}R_{NS}$ and QPO frequencies of order $\nu_2 \times 10^2$ hz,  velocity of crustal oscillations is  $\sim 2\pi \times 10^{6}A_{-2}\nu_2$ cm/s at the base of the crust. This constitutes  $\sim 10^{45}$ ergs in energy of oscillation.

The velocity differential between the crust and neutron star
superfluid associated with the toroidal oscillations is, even in a conservative estimate of Timokhin et al (2007)(pair multiplicities $\eta$ of order 100, $A_{-2}\sim 1/\eta \sim 10^{-2}$),   larger than the estimated critical
velocity difference between crust and the superfluid. The Vela pulsar, for example, which has an age comparable to that of magnetars, has a  critical
angular velocity difference between crust and the superfluid that has been  estimated (Alpar and Baykal, 1994) to be in the range $0.0049 - 0.013$rad/s,
see also Lyne et al.(2000) whose  numbers imply $5.5\times 10^{-3}$rad/s. These estimates assume that the critical velocity has been marginally reached at the time of the glitch and thus may be conservative. So for the Vela pulsar the critical crust-superfluid velocity
lag is in the range $5\times10^{3}-10^{4}$cm/s.  This translates to an energy in crustal motion of less than $10^{40}$ ergs to set off a glitch, of which most may go into the core either via magnetic coupling (Levin, 2007) or, if damped locally,  via heat conduction (e.g. Eichler and Cheng 1989, Lyubarsky, Eichler and Thompson, 2002)). Thus, it is entirely possible for an unpinning event to occur without setting off a significant X-ray brightening. This is consistent with the observation that, while all radiative enhancements are indicated by the present data set to be  associated with glitches or very similar timing irregularities,  not all glitches are associated with radiative enhancements (DKG 2009).

If magnetar cores are subject to cooling by the direct Urca process, then they may be cooler than the Vela pulsar at the bottom of their crusts, so we estimate the maximum lag velocity for cold NS  crustal material.  The Magnus force on a unit
length of vortex line can be written as
\begin{equation}
\mathbf{f}_{M}=\rho _{s} \,  \bm{\kappa} \times \left( \mathbf{v}_{L}-\mathbf{v}%
_{s}\right)
\end{equation}
where $\bm{\kappa}$ is aligned along the vortex and its magnitude
$\kappa=h/2m_{n}$, $\mathbf{v}_{L}$ is the velocity of the
line in the laboratory system, $\mathbf{v}_{s}$ is the superflow
velocity and $\rho_{s}$ is the superfluid density. When $f_M$ exceeds the maximum pinning force $f_p \equiv
10^{16}f_{16} $ dyn/cm, the vortices unpin even without a
significant temperature rise. From equation (1) above, we can see
that a velocity difference greater than
\begin{equation}
\delta v=\frac{f_{p}}{\rho _{s} \kappa} \approx 5\times 10^{5} \,
\frac{f_{16}}{\rho _{13}} \, \mathrm{\frac{cm}{s}}
\end{equation}
triggers vortex unpinning. Here  $\rho _{s}=\rho _{13}\ 10^{13}
\, \mathrm{g/cm^{3}}$.
At present time there is no consensus about the value of
the maximum pinning force $f_p$. Several approaches
give different values for pinning and binding energies for vortices
in neutron star (see e.g. Jones 1997, 1998; Donati \& Pizzochero 2003, 2004, 2006; Avogadro et al. 2007).
Link and Cutler (2002) suggested
 that $10^{15}\leq f_{p} \leq 2\times 10^{16}$  dyn/cm.
Thus,
the velocity of crustal seismic oscillations  $2\pi \, A \,
 \nu_{osc}$ is large enough to cause vortex unpinning inside
the inner crust for amplitudes A of $10^{-2}R_{NS}$, and, if $f_{p} = 10^{15}$  dyn/cm,  large
enough even for amplitudes of $\sim 10^{-4}R_{NS}$, and frequencies of order
$10^2$ Hz. Acoustic modes, which have considerably higher frequencies
than toroidal modes, could trigger unpinning with even lower
amplitudes.

\section{Summary and Further Discussion}

QPOs in magnetars are an unprecedented observation of crustal motion  relative to superfluid, $v=2\pi[A/10^{-2} R_{NS}][\nu_{QPO}/100Hz)][R_{NS}/10^{6}cm]10^6$cm/s.  By our estimate, this relative velocity may greatly exceed the velocity difference that vortex pinning can support, which is estimated to be about $10^{4}$ cm/s for the Vela pulsar, and $10^4$ to $10^5$ cm/s for a cold crust.
We have suggested that brightening events are triggered by mechanical energy release that generates enough relative motion between crust and superfluid to unpin the vortices. Although only a small fraction of the total radiated X-ray energy need go into such motion in order to unpin the vortices, consistent with the observation that many glitches are unaccompanied by noticeable radiative enhancements, the fact that all such radiative enhancements appear to be accompanied by timing irregularities suggests that they may also be accompanied by torsional oscillations.
This picture would be tested  by better pinpointing  the times of glitches and brightening onsets. The glitch should coincide or precede the first sign of X-ray brightening.  By contrast, observing the glitch only after the X-ray brightening event was well under way would favor a thermal unpinning mechanism.

Detecting QPO behavior coincident with the glitch would confirm the present hypothesis. However, because QPO features last only several minutes following a catastrophic energy release, it would be necessary to "catch" the glitch while it is in progress, and it is not clear that this is likely with present detectors, unless it is accompanied by $\gamma$-rays and/or a large photon flux.

As noted by Timokhin, Eichler and Lyubarsky (2007), torsional oscillations enhance magnetospheric currents and the extent of resonant cyclotron upscattering of thermal X-rays, and the upscattered photons would appear as pulsed radiation above 2 KeV. An increase of even only $\sim 0.03$ in the cyclotron optical depth could be comparable to the steady value.  Thus, torsional oscillation could cause impulsive hard X-ray enhancement.

AXP's, in fact show impulsive X-ray bursts that last only $10\sim 10^2$s (Gavriil, Dib, and Kaspi, 2009), and typically contain less than $10^{38}$ergs. The timescale is too long to be the magnetospheric dynamical timescale, and too short to be the cooling of a pair supported atmosphere (Eichler and Cheng, 1989: Eichler et al. 2003). However, it agrees with the timescale for QPO behavior. The energetics are low enough that a glitch is not necessarily triggered, but the nonthermal X-ray burst might be a necessary consequence of sufficiently large amplitude toroidal oscillation. Such events should in principle be analyzed for QPO components, though the low luminosities render positive detection questionable.

Increases of magnetospheric currents might increase coherent  microwave and IR emission (Eichler, Gedalin and Lyubarsky 2002).Such IR transients of peak flux densities $\gg$ 1 mJy  might be detectable with the Planck satellite or a similarly designed instrument, if it were to monitor the Galactic plane. A dedicated small telescope could monitor magnetars for transients in the near IR and optical.

Incomplete glitch recoveries are sometimes reported for magnetars (DKG 2008) and they can be attributed to a long lasting temperature rise in the lower crust, which allows less velocity difference between
the crust and the neutron superfluid.  However, some of what used to
be known as "persistent" emission of magnetars is in fact long term,
but transient, afterglow, as evidenced by the slow decline of the
steady X-ray emission from SGR 1900+14. This can be accounted for by
long term cooling of the lower crust by heat conduction into a
cooler core (Eichler et al. 2006). This suggests that incomplete glitch recovery could complete itself over a much longer timescale if given time to do so before the next glitch. It would therefore be interesting
to monitor incomplete glitch recoveries over a timescale of
years to look for gradual, long term recovery over the cooling timescale of
the lower crust, given the accompanying assumption of a cool core.
Ultimately, this might provide further constraints on the nature of
episodic heating in the lower crust.

The authors thank Drs.  Y. Lyubarsky, V. Kaspi, and A. Cumming for
very helpful conversations. DE acknowledges the hospitality of the McGill Department of Physics, where this work began. Support from the Israel Science
Foundation,  the Israel-U.S. Binational Science Foundation, and the
Joan and Robert Arnow Chair of Theoretical Astrophysics  is
also acknowledged.

\end{document}